\documentclass[12pt]{article}
\usepackage{a4wide}
\usepackage{latexsym}
\usepackage{cite}

\usepackage{pslatex}
\usepackage[latin1]{inputenc}
\usepackage[T1]{fontenc}

\def\bq{\begin{eqnarray}}
\def\eq{\end{eqnarray}}
\def\l{\langle}
\def\r{\rangle}

\begin{document}

\thispagestyle{empty}

\begin{flushright}
  MZ-TH/06-15
\end{flushright}

\vspace{1.5cm}

\begin{center}
  {\Large\bf The forward-backward asymmetry at NNLO revisited\\
  }
  \vspace{1cm}
  {\large Stefan Weinzierl\\
\vspace{2mm}
      {\small \em Institut f{\"u}r Physik, Universit{\"a}t Mainz,}\\
      {\small \em D - 55099 Mainz, Germany}\\
  } 
\end{center}

\vspace{2cm}

\begin{abstract}\noindent
  {
I reconsider the forward-backward asymmetry for flavoured quarks in electron-positron
annihilation.
I suggest an infrared-safe definition of this observable, such that the asymmetry may be computed
in perturbative QCD with massless quarks.
With this definition, the first and second order QCD corrections are computed.
   }
\end{abstract}

\vspace*{\fill}

\newpage

\section{Introduction}
\label{sec:intro}

The forward-backward asymmetry of bottom and charm production in electron-positron annihilation is a key observable
for precision measurements.
It provides a precise determination of the effective weak mixing angle $\sin^2\theta^{lept}_{W,eff}$,
which in turn leads to constraints on the mass of the yet unobserved Higgs boson.
In a recent report of the electroweak working group \cite{unknown:2005em} on precision measurements on the
$Z$ resonance, the forward-backward asymmetry for $b$-quarks shows the largest discrepancy with
the Standard Model predictions of about $2.8$ standard deviations.
This motivates a closer re-examination of the forward-backward asymmetries.

The forward-backward asymmetries for charm and bottom quarks are measured experimentally
with a precision at the per cent level.
To match this experimental precision 
the inclusion of QCD corrections in a theoretical calculation is mandatory.
The next-to-leading order (NLO) corrections to the forward-backward asymmetry have been calculated in
\cite{Jersak:1979uv,Korner:1985dt,Arbuzov:1991pr,Djouadi:1994wt,Lampe:1996rt}.
The next-to-next-to-leading order (NNLO) corrections have been considered by three groups
\cite{Altarelli:1992fs,Ravindran:1998jw,Catani:1999nf} in massless QCD.
The last calculation by Catani and Seymour \cite{Catani:1999nf}
was performed to clarify the disagreement between the first two calculations.
Furthermore these authors pointed out, that the forward-backward asymmetry is not
infrared-safe, if the direction is defined by the direction of the flavoured quark, as it was
done in the first two calculations.
Nor is the quantity infrared-safe, if the direction is defined by the thrust axis, a definition which has been
used in the experimental analysis \cite{Abbaneo:1998xt}.
For a fully massive calculation only partial results for the NNLO corrections are available at the moment
\cite{Bernreuther:2000zx,Bernreuther:2006vp}.

In order to reduce uncertainties related to non-perturbative effects, an infrared-safe definition 
of the forward-backward asymmetry
is highly desirable.
In this letter I reconsider the forward-backward asymmetry for flavoured quarks.
The first purpose is to suggest a definition of the forward-backward asymmetry, which
is infrared-safe. This definition is a direct application of the infrared-safe definition
of a heavy-quark jet, introduced in \cite{Banfi:2006hf}.
An infrared-safe observable can be calculated reliably in perturbation theory.
The second purpose of this letter is to compute for the infrared-safe definition
of the forward-backward asymmetry the NLO and NNLO corrections.
It is well known that the genuine two-parton contributions to
the corrections cancel in the asymmetry.
Therefore the NNLO corrections could be obtained by a dedicated NLO calculation,
a fact already noted in ref.~\cite{Altarelli:1992fs}.
In this paper a different and more direct approach is taken.
The corrections are obtained with the help of the recently developed 
general program for NNLO corrections to the
process $e^+ e^- \rightarrow \mbox{2 jets}$
\cite{Weinzierl:2006ij}.
It turns out that these corrections are small, making the observable
suggested here a candidate for high precision measurements at a future linear collider.

This article is organised as follows:
Basic definitions related to the forward-backward asymmetry are reviewed in the next section
and an infrared-safe definition of this observable is presented.
In sect.~\ref{sect:outline} a few technical details are outlined for the calculation of the 
NNLO corrections.
The numerical results are presented in sect.~\ref{sect:results}.
Finally, sect.~\ref{sect:conclusion} contains the conclusions.

\section{Definitions}
\label{sec:def}

The forward-backward asymmetry is defined by
\bq
\label{def1}
 A_{FB} & = & \frac{\sigma_A}{\sigma_S},
\eq
where $\sigma_S$ and $\sigma_A$ are the symmetric and antisymmetric cross sections. These are expressed
in terms of the forward and backward cross sections $\sigma_F$ and $\sigma_B$ as follows:
\bq
\label{def2}
 \sigma_S = \sigma_F + \sigma_B,
 & &
 \sigma_A = \sigma_F - \sigma_B.
\eq
The definition of the forward and backward cross sections $\sigma_F$ and $\sigma_B$ requires a little bit more 
care.
Loosely speaking, we think about the forward cross section as 
that part of the total cross section in which the flavoured quark
is observed in the forward hemisphere, and correspondingly we think
about the backward cross section as that part in which the flavoured quark
is observed in the backward hemisphere.
One is therefore tempted to define the direction entering the definitions of
$\sigma_F$ and $\sigma_B$ as the direction of the flavoured quark.
This definition was used in refs.~\cite{Altarelli:1992fs,Ravindran:1998jw,Catani:1999nf}.
An alternative definition defines the direction by the thrust axis, the two-fold
ambiguity inherent in this definition is resolved by singling out the direction which is closer to the 
flavoured quark.
Both definitions are not infrared-safe, a fact which was pointed out in ref.~\cite{Catani:1999nf}.
This failure of infrared-safeness is related to the $q \rightarrow q q \bar{q}$ splitting function,
which occurs in the NNLO corrections for the first time.

To cure the situation requires a more careful definition of the direction
entering the determination of $\sigma_F$ and $\sigma_B$.
It should be taken as the jet axis of a flavoured quark jet.
Banfi, Salam and Zanderighi \cite{Banfi:2006hf}
recently proposed an infrared-safe definition for 
a flavoured quark jet.
Starting from the DURHAM algorithm \cite{Stirling:1991ds}, 
they modified the resolution criteria of the DURHAM algorithm
\bq
 y_{ij}^{DURHAM} & = & 
 \frac{2(1-\cos\theta_{ij})}{Q^2} 
 \mbox{min}(E_i^2,E_j^2)
\eq
towards
\bq
 y_{ij}^{flavour} & = & 
 \frac{2(1-\cos\theta_{ij})}{Q^2} \times \left\{ 
 \begin{array}{ll}
 \mbox{min}(E_i^2,E_j^2), & \mbox{softer of $i$, $j$ is flavourless}, \\
 \mbox{max}(E_i^2,E_j^2), & \mbox{softer of $i$, $j$ is flavoured}. \\
 \end{array} \right.
\eq
The resulting jet algorithm is called the 
flavour-$k_\perp$ algorithm.
Each pseudo-particle carries an additional additive quantum number, which is the flavour number.
If two pseudo-particles are merged, the flavour numbers are added.
The original DURHAM algorithm is infrared unsafe for flavoured jets at order ${\cal O}(\alpha_s^2)$,
which can be seen by considering a large-angle soft gluon which in turns splits into a flavoured
$q \bar{q}$ pair.
The flavour-$k_\perp$ algorithm ensures that these $q \bar{q}$ pairs are merged together first, by not introducing
spurious closeness to other particles in the resolution variable.

As a concrete example let us consider the calculation of the forward-backward asymmetry for $b$-quarks.
We associate the flavour number $+1$ to a $b$-quark and the flavour number $-1$ to a $\bar{b}$-quark.
All other particles (gluons and $d$-, $u$-, $s$- and $c$-quarks) are assigned the flavour number $0$.
If we combine two $b$-quarks, the resulting pseudo-particle has flavour number $2$, whereas if we combine
a $b$-quark with an $\bar{b}$-quark, the resulting pseudo-particle has flavour number $0$.
A pseudo-particle with a flavour number equal to zero is called flavourless, otherwise it is called
flavoured.
For the calculation of the forward-backward asymmetry for $c$-quarks, the r\^oles of the $b$- and $c$-quark
are exchanged.

This brings us to the following definition of the forward-backward asymmetry for flavoured quark jets:
We first select two-jet events according to the flavour-$k_\perp$ algorithm and a specified resolution parameter
$y_{cut}$.
The jet axis defines if a jet lies in the forward or backward hemisphere.
For the forward cross section we require a jet with flavour number greater than zero in
the forward hemisphere.
The backward cross section is defined similar: Here we require a jet with flavour number greater than zero 
in the backward hemisphere.
With these definitions of the forward and backward cross sections, the forward-backward asymmetry is then
given by eq.~(\ref{def1}) and eq.~(\ref{def2}).

This definition is infrared-safe and all quantities can be computed in perturbation theory.
In particular we are interested in the value of the forward-backward asymmetry on the $Z$-resonance.
Here the cross section is dominated by the $Z$-boson exchange.
If we neglect the photon exchange, the leading order result for the 
forward-backward asymmetry on the $Z$-resonance $s=m_Z^2$ is given by
\bq
 A_{FB}^{(0)} & = & \frac{3}{4} {\cal A}_e {\cal A}_q,
\eq
where
\bq 
 {\cal A}_f & = & \frac{2 v_f a_f}{v_f^2 + a_f^2}.
\eq
The quantities $v_f$ and $a_f$ are 
the vector and axial-vector couplings of a fermion to the $Z$-boson and given by
\bq
 v_f = I_3 - 2 Q \sin^2 \theta_W,
 & &
 a_f = I_3.
\eq

\section{Outline of the calculation}
\label{sect:outline}

In this section I give a brief outline of the calculation of the NLO and the NNLO QCD corrections
to the forward-backward asymmetry.
This calculation is based on a general numerical program for NNLO (and NLO) corrections 
to the process $e^+ e^- \rightarrow \mbox{2 jets}$
\cite{Weinzierl:2006ij}.
To this order the following amplitudes enter:
The amplitudes for $e^+ e^- \rightarrow q \bar{q}$ up to two-loops, the amplitudes
$e^+ e^- \rightarrow q g \bar{q}$ up to one-loop, as well as the Born amplitudes
for $e^+ e^- \rightarrow q g g \bar{q}$ and $e^+ e^- \rightarrow q \bar{q} q' \bar{q}'$.
As in ref.~\cite{Catani:1999nf} closed triangle diagrams in the ${\cal O}(\alpha_s^2)$-part are neglected.
These diagrams are expected to give numerically a very small contribution.
The program uses the subtraction method to 
handle infrared divergences 
\cite{Weinzierl:2003fx,Gehrmann-DeRidder:2005cm}.
For the calculation of the forward-backward asymmetry information on the flavour of the final-state
partons had to be added.
This is straightforward for amplitudes involving only one pair of quarks, e.g. for the amplitudes
$e^+ e^- \rightarrow q \bar{q} + n \;\mbox{gluons}$.
Here I would like to discuss the contribution from the four-quark final-state.
The colour decomposition of the tree amplitude
for $e^+ e^- \rightarrow q \bar{q} q' \bar{q}'$ is
\bq
{\cal A}_{4,q\bar{q}q'\bar{q}'}^{(0)}(q_1,\bar{q}_2,q_3',\bar{q}_4',e^+_5,e^-_6)
 & = &
 e^2 g^2
 \left\{ 
        \frac{1}{2} \left( \delta_{14} \delta_{32} - \frac{1}{N} \delta_{12} \delta_{34} \right) 
        \chi^{(0)}_4(1,2,3,4) 
 \right.
 \nonumber \\
 & & \left.
        - \delta_{flav} 
        \frac{1}{2} \left( \delta_{12} \delta_{34} - \frac{1}{N} \delta_{14} \delta_{32} \right) 
        \chi^{(0)}_4(1,4,3,2) \right\}, 
 \nonumber \\
\chi^{(0)}_4(1,2,3,4) 
 & = & 
 c_0(1) A^{(0)}(1,2,3,4) + c_0(3) A^{(0}(3,4,1,2).
\eq
In this formula $\delta_{ij}$ denotes a Kronecker delta in colour space. $\delta_{flav}$
equals one, if the flavours of the quarks $q$ and $q'$ are identical and zero otherwise.
The variable 
$c_0(q)$ denotes the electroweak coupling of the quark $q$ and is given by
\bq
c_0(q) = -Q^q + v^e v^q {\cal P}_Z(s), 
 & &
{\cal P}_Z(s) = \frac{s}{s-M_Z^2+ i \Gamma_Z M_Z}.
\eq
$v^e$ and $v^q$ are the couplings of the $Z$-boson to an electron and a quark.
$A^{(0)}(1,2,3,4)$ is the colour-ordered partial amplitude. Explicit expressions
for these amplitudes can be found for example in \cite{Bern:1997ka}.
In squaring the amplitude it is convenient to split the resulting expression into
a leading-colour piece and a sub-leading-colour piece and to distinguish the cases 
where the two pairs of quark are identical or not.
We therefore write for the squared amplitude, summed over colours, helicities and flavours:
\bq
\left| {\cal A}^{(0)}_{4,q\bar{q}q'\bar{q}'} \right|^2
 & = & 
 {\cal M}(1,2,3,4) 
 =
 \left. {\cal M}(1,2,3,4) \right|_{lc,{\footnotesize \mbox{not id}}}
 +
 \left. {\cal M}(1,2,3,4) \right|_{lc,{\footnotesize \mbox{id}}}
 +
 \left. {\cal M}(1,2,3,4) \right|_{sc,{\footnotesize \mbox{id}}}.
 \nonumber \\
\eq
Note that there is no sub-leading-colour contribution for non-identical quarks.
The individual contributions are given by
\bq
 \left. {\cal M}(1_Q,2_{\bar Q},3,4) \right|_{lc,{\footnotesize \mbox{not id}}}
 & = &
e^4 g^4 \frac{(N^2 -1)}{4} |\chi^{(0)}_4(1,2,3,4)|^2,
 \nonumber \\
 \left. {\cal M}(1_Q,2_{\bar Q},3_Q,4_{\bar Q}) \right|_{lc,{\footnotesize \mbox{id}}}
 & = &
e^4 g^4 \frac{(N^2 -1)}{4} 
         \left( |\chi^{(0)}_4(1,2,3,4)|^2 + |\chi^{(0)}_4(1,4,3,2)|^2 \right),
 \nonumber \\
 \left. {\cal M}(1_Q,2_{\bar Q},3_Q,4_{\bar Q}) \right|_{sc,{\footnotesize \mbox{id}}}
 & = &
e^4 g^4 \frac{(N^2 -1)}{4 N} \; 2 \;\mbox{Re}
         \left[ \chi^{(0)}_4(1,2,3,4)^\ast \chi^{(0)}_4(1,4,3,2) \right].
\eq
To render this four-particle final-state contribution integrable, we subtract
approximation terms, which are re-added to the three- and two-parton final-state
terms.
In general the double-un\-re\-solved contribution is given by
\bq
\l {\cal O} \r^{NNLO}_{n+2} & = &
 \int \left( {\cal O}_{n+2} \; d\sigma_{n+2}^{(0)} 
             - {\cal O}_{n+1} \circ d\alpha^{(0,1)}_{n+1}
             - {\cal O}_{n} \circ d\alpha^{(0,2)}_{(0,0),n} 
             + {\cal O}_{n} \circ d\alpha^{(0,2)}_{(0,1),n}
      \right).
\eq
The notation ${\cal O} \circ d\alpha$ is a reminder, that
in general the approximation is a sum of terms
\bq
{\cal O} \circ d\alpha & = & \sum {\cal O} \; d\alpha
\eq
and the mappings used to relate the $n+2$ or $n+1$ parton configuration to a $n+1$ or $n$ parton configuration
differs in general for each summand.
The individual subtraction terms read as follows:
The NLO subtraction terms are given by
\bq
 \left. d\alpha_{3,q\bar{q}q'\bar{q}'}^{(0,1)} \right|_{lc,{\footnotesize \mbox{not id}}}
 & = &
 \left(N_f-1\right) 
  \frac{N}{2} \left[ E_3^0(1,3,4) + E_3^0(2,4,3) \right]
 \circ \left| {\cal A}_3^{(0)} \right|^2,
 \nonumber \\
 \left. d\alpha_{3,q\bar{q}q'\bar{q}'}^{(0,1)} \right|_{lc,{\footnotesize \mbox{id}}} 
 & = &
 \frac{1}{4} 
  \frac{N}{2} \left[ E_3^0(1,3,4) + E_3^0(2,4,3) + E_3^0(3,1,2) + E_3^0(4,2,1) 
 \right. 
 \nonumber \\
 & &
 \left.
   + E_3^0(1,3,2) + E_3^0(2,4,1) + E_3^0(3,1,4) + E_3^0(4,2,3) \right]
 \circ \left| {\cal A}_3^{(0)} \right|^2,
 \nonumber \\
 \left. d\alpha_{3,q\bar{q}q'\bar{q}'}^{(0,1)} \right|_{sc,{\footnotesize \mbox{id}}}
 & = & 0.
\eq
Here, ${\cal A}_3^{(0)}$ denotes the Born amplitude for $e^+ e^- \rightarrow q g \bar{q}$
and $E_3^0$ denotes the $q q \bar{q}$-antenna function, which can be found
in \cite{Gehrmann-DeRidder:2005cm}.
The subtraction terms for double unresolved contributions read
\bq
\lefteqn{
 \left. d\alpha^{(0,2)}_{(0,0),2,q\bar{q}q'\bar{q}'} \right|_{lc,{\footnotesize \mbox{not id}}}
 = 
 \left(N_f-1\right) \frac{C_F}{2} 
                        B_4^0(2,4,3,1)
 \circ \left| {\cal A}_2^{(0)} \right|^2 } & &,
 \nonumber \\
\lefteqn{
 \left. d\alpha^{(0,2)}_{(0,0),2,q\bar{q}q'\bar{q}'} \right|_{lc,{\footnotesize \mbox{id}}}
 = } & &
 \nonumber \\
 & &
 \frac{1}{4} \frac{C_F}{2} \left[  
                        B_4^0(2,4,3,1) + B_4^0(4,2,1,3) 
                        + B_4^0(2,4,1,3) + B_4^0(4,2,3,1) \right]
 \circ \left| {\cal A}_2^{(0)} \right|^2, 
 \nonumber \\
\lefteqn{
 \left. d\alpha^{(0,2)}_{(0,0),2,q\bar{q}q'\bar{q}'} \right|_{sc,{\footnotesize \mbox{id}}}
 = } & &
 \nonumber \\
 & &
 - \frac{C_F}{2N} \left[ C_4^0(2,4,3,1) + C_4^0(4,2,1,3) + C_4^0(2,4,1,3) + C_4^0(4,2,3,1) \right]
 \circ \left| {\cal A}_2^{(0)} \right|^2.
\eq
${\cal A}_2^{(0)}$ denotes the Born amplitude for $e^+ e^- \rightarrow q \bar{q}$.
$B_4^0(q,q',\bar{q}',\bar{q})$ and $C_4^0(q,q,\bar{q},\bar{q})$ denote the
four-quark double-unresolved antenna functions. Explicit expressions can be found
in \cite{Weinzierl:2006ij}.
Finally, the iterated subtraction terms for double unresolved contributions read
\bq
\lefteqn{
 \left. d\alpha^{(0,2)}_{(0,1),2,q\bar{q}q'\bar{q}'} \right|_{lc,{\footnotesize \mbox{not id}}}
 = }
 \nonumber \\
 & &
 \left(N_f-1\right) 
  \frac{N}{2} \left[ E_3^0(1,3,4) + E_3^0(2,4,3) \right]
 \circ \frac{N^2-1}{2N}
 A_3^0(1',2',3') \left| {\cal A}_2^{(0)}(1'', 2'') \right|^2,
 \nonumber \\
\lefteqn{
 \left. d\alpha^{(0,2)}_{(0,1),2,q\bar{q}q'\bar{q}'} \right|_{lc,{\footnotesize \mbox{id}}}
 = 
 \frac{1}{4} 
  \frac{N}{2} \left[ E_3^0(1,3,4) + E_3^0(2,4,3) + E_3^0(3,1,2) + E_3^0(4,2,1) 
 + E_3^0(1,3,2) 
 \right. 
}
 \nonumber \\
 & &
 \left.
 + E_3^0(2,4,1) + E_3^0(3,1,4) + E_3^0(4,2,3) \right]
 \circ \frac{N^2-1}{2N}
 A_3^0(1',2',3') \left| {\cal A}_2^{(0)}(1'', 2'') \right|^2,
 \hspace*{20mm}
 \nonumber \\
\lefteqn{
 \left. d\alpha^{(0,2)}_{(0,1),2,q\bar{q}q'\bar{q}'} \right|_{sc,{\footnotesize \mbox{id}}}
 = 0. }
\eq
$A_3^0$ is the NLO $q g \bar{q}$ antenna function and can be found in \cite{Gehrmann-DeRidder:2005cm}.

\section{Numerical results}
\label{sect:results}

In this section I give numerical results for the forward-backward asymmetry both for
$b$-quarks and $c$-quarks.
The nominal choice of input parameters is $N=3$ colours and
$N_f =5$ massless quarks.
I take the electromagnetic coupling to be $\alpha(m_Z) = 1/127.9$ and the strong
coupling to be $\alpha_s(m_Z) = 0.118$. The numerical values of the $Z^0$-mass and width
are $m_Z = 91.187$ GeV and $\Gamma_Z = 2.490$ GeV. For the weak mixing angle I use
$\sin^2 \theta_W = 0.230$.
The center of mass energy is $\sqrt{s} = m_Z$ and 
the renormalization scale is set equal to $\mu^2 = s$.

We select two-jet events, where the jets are defined by the
flavour-$k_\perp$ algorithm.
The recombination prescription is given by the E-scheme.
We consider jets with flavour number greater than zero.
The direction of a jet is given by the jet axis. 
 
We expand the symmetric cross section $\sigma_S$, the asymmetric cross section $\sigma_A$, the forward cross
section $\sigma_F$ and the backward cross section $\sigma_B$ in powers of $\alpha_s$:
\bq
 \sigma_K & = & \sigma_K^{(0)}
 \left( 1 + \frac{\alpha_s}{2\pi} B_K + \left( \frac{\alpha_s}{2\pi} \right)^2 C_K
 \right)
 + {\cal O}\left(\alpha_s^3\right), \;\;\;\mbox{where}\; K \in \{S,A,F,B\}.
\eq
Then the forward-backward asymmetry is given by
\bq
 A_{FB} & = & A_{FB}^{(0)} 
 \left( 1 + \frac{\alpha_s}{2\pi} B_{FB}
   + \left( \frac{\alpha_s}{2\pi} \right)^2 C_{FB} 
 \right) + {\cal O}\left(\alpha_s^3\right),
\eq
where the leading order result is given by
\bq
 A_{FB}^{(0)} & = & \frac{\sigma_A^{(0)}}{\sigma_S^{(0)}}
\eq
and the coefficients $B_{FB}$ and $C_{FB}$ can be expressed as follows:
\bq
 B_{FB} & = & B_A - B_S,
 \nonumber \\
 C_{FB} & = & C_A - C_S - \left( B_A - B_S \right) B_S.
\eq
The NNLO coefficients can be decomposed into independent colour structures:
\bq
 C_K & = & \frac{1}{4} \left( N^2-1 \right) \left( C_{K,lc} + \frac{1}{N} C_{K,nf} + \frac{1}{N^2} C_{K,sc} \right),
 \;\;\;\mbox{where}\; K \in \{S,A,F,B,FB\}.
\eq
\begin{table}
\begin{center}
\begin{tabular}{|l|ll|}
\hline
 $y_{cut}$ & $B_{FB,b}$ & $C_{FB,b}$ \\
\hline
 $0.01$ & $-0.070 \pm 0.005$ & $-0.4 \pm 0.8$ \\
 $0.03$ & $-0.145 \pm 0.003$ & $-1.7 \pm 0.5$ \\
 $0.1$  & $-0.294 \pm 0.002$ & $-4.3 \pm 0.3$ \\ 
 $0.3$  & $-0.512 \pm 0.001$ & $-10.2 \pm 0.1$ \\
 $0.9$  & $-0.565 \pm 0.001$ & $-13.4 \pm 0.1$ \\
\hline
\end{tabular}
\caption{\label{resNNLO_b}
QCD corrections to the forward-backward asymmetry of $b$-quarks for various values of the jet resolution parameter $y_{cut}$.
The NLO correction is given by $B_{FB}$, while the NNLO correction is given by $C_{FB}$.
}
\end{center}
\end{table}
\begin{table}
\begin{center}
\begin{tabular}{|l|ll|}
\hline
 $y_{cut}$ & $B_{FB,c}$ & $C_{FB,c}$ \\
\hline
 $0.01$ & $-0.070 \pm 0.005$ & $-0.5 \pm 0.7$ \\
 $0.03$ & $-0.145 \pm 0.003$ & $-2.1 \pm 0.5$ \\
 $0.1$  & $-0.294 \pm 0.002$ & $-4.8 \pm 0.2$ \\ 
 $0.3$  & $-0.513 \pm 0.001$ & $-12.1 \pm 0.2$ \\
 $0.9$  & $-0.565 \pm 0.001$ & $-15.9 \pm 0.1$ \\
\hline
\end{tabular}
\caption{\label{resNNLO_c}
QCD corrections to the forward-backward asymmetry of $c$-quarks for various values of the jet resolution parameter $y_{cut}$.
The NLO correction is given by $B_{FB}$, while the NNLO correction is given by $C_{FB}$.
}
\end{center}
\end{table}
\begin{table}
\begin{center}
\begin{tabular}{|l|rrr|}
\hline
 $y_{cut}$ & $\frac{1}{4} (N^2-1) C_{FB,b,lc}$ & $\frac{1}{4 N} (N^2-1) C_{FB,b,nf}$ & $\frac{1}{4 N^2} (N^2-1) C_{FB,b,sc}$ \\
\hline
 $0.01$ & $-1.3 \pm 0.8$ & $0.85 \pm 0.11$ & $-0.20 \pm 0.09$ \\
 $0.03$ & $-3.3 \pm 0.5 $ & $1.39 \pm 0.06$ & $-0.24 \pm 0.05$ \\
 $0.1$  & $-6.6 \pm 0.3$ & $2.07 \pm 0.04$ & $-0.06 \pm 0.02$ \\
 $0.3$  & $-11.4 \pm 0.1$ & $0.97 \pm 0.02$ & $0.21 \pm 0.01$ \\
 $0.9$  & $-12.7 \pm 0.1$ & $-0.96 \pm 0.02$ & $0.32 \pm 0.01$ \\
\hline
\end{tabular}
\caption{\label{rescolourNNLO_b}
The results for the NNLO correction to the forward-backward asymmetry of $b$-quarks for various values 
of the jet resolution parameter $y_{cut}$ split into the different colour structures.
}
\end{center}
\end{table}
\begin{table}
\begin{center}
\begin{tabular}{|l|rrr|}
\hline
 $y_{cut}$ & $\frac{1}{4} (N^2-1) C_{FB,c,lc}$ & $\frac{1}{4 N} (N^2-1) C_{FB,c,nf}$ & $\frac{1}{4 N^2} (N^2-1) C_{FB,c,sc}$ \\
\hline
 $0.01$ & $-1.3 \pm 0.7$ & $0.77 \pm 0.10$ & $-0.17 \pm 0.08$ \\
 $0.03$ & $-3.5 \pm 0.5$ & $1.26 \pm 0.06$ & $-0.26 \pm 0.06$ \\
 $0.1$  & $-6.7 \pm 0.2$ & $1.67 \pm 0.03$ & $-0.09 \pm 0.03$ \\
 $0.3$  & $-11.7 \pm 0.2$ & $-0.66 \pm 0.03$ & $0.21 \pm 0.01$ \\
 $0.9$  & $-12.6 \pm 0.1$ & $-3.53 \pm 0.03$ & $0.32 \pm 0.01$ \\
\hline
\end{tabular}
\caption{\label{rescolourNNLO_c}
The results for the NNLO correction to the forward-backward asymmetry of $c$-quarks for various values 
of the jet resolution parameter $y_{cut}$ split into the different colour structures.
}
\end{center}
\end{table}
The leading order result for the bottom and charm quark (including the photon contribution) is given by
\bq
 A_{FB,b}^{(0)} = 0.11161,
 & &
 A_{FB,c}^{(0)} = 0.08003.
\eq
The leading order result is independent of the jet resolution parameter $y_{cut}$.
The results for the QCD corrections to the forward-backward asymmetry of $b$-quarks
for various values of the jet resolution parameter $y_{cut}$ can be found in table~\ref{resNNLO_b}.
The corresponding results for the QCD corrections to the forward-backward asymmetry of $c$-quarks
are given in table~\ref{resNNLO_c}.
The errors are those of the Monte-Carlo integration used in the numerical program and can be reduced by
increasing the number of evaluations of the integrand.
Tables~\ref{rescolourNNLO_b} and \ref{rescolourNNLO_c} give the NNLO results for the individual colour structures.

The QCD corrections are small. By comparing the NLO corrections with the ones reported in ref.~\cite{Catani:1999nf}
for the quark axis and thrust axis definition, one observes that the corrections to the forward-backward asymmetry
defined by the jet axis are for all values of $y_{cut}$ smaller than the ones reported in ref.~\cite{Catani:1999nf}.
In addition one deduces from the dependence on $y_{cut}$ that the corrections are further reduced by enforcing
more stringent two-jet cuts.
In addition the definition based on the jet axis is by construction infrared-safe, as opposed to the definitions
by the quark axis or the thrust axis.
This makes the forward-backward asymmetry defined by the jet axis a suitable precision observable, whose
QCD corrections are well under control.
Therefore this definition is a candidate for high precision studies at a future linear collider.

As a last remark it is worth pointing out that the numerical program used for this letter can also be used
with slight modifications to obtain the corresponding QCD corrections to the left-right asymmetry and the combined
left-right forward-backward asymmetry.

\section{Conclusions}
\label{sect:conclusion}

In this article I reconsidered the forward-backward asymmetry for flavoured quarks
in electron-positron annihilation with a two-fold purpose:
The first point was to provide an infrared-safe definition of this quantity.
The definition is based on the concept of a flavoured quark jet, where the direction is defined
by the jet axis.
With an infrared-safe definition at hand, the observable can be computed perturbatively 
in massless QCD.
In the second part of this article I provided results for the NLO and NNLO corrections to this
observable.
These corrections are small, in particular if stringent two-jet cuts are enforced.
Therefore this definition of the forward-backward asymmetry is an ideal candidate 
for precision studies.


\end{document}